\documentclass[fleqn,10pt]{wlscirep}
\usepackage[utf8]{inputenc}
\usepackage[T1]{fontenc}
\usepackage{graphicx,calc}
\usepackage{todonotes}
\usepackage[nameinlink]{cleveref}
\usepackage{multirow}
\usepackage{blindtext}
\usepackage[most]{tcolorbox}
\usepackage{graphicx}
\usepackage{xcolor}
\usepackage{mathtools}
\newlength\myheight
\newlength\mydepth
\settototalheight\myheight{Xygp}
\definecolor{myred}{RGB}{117, 2, 27}
\definecolor{myyellow}{RGB}{194, 175, 2}
\definecolor{mygreen}{RGB}{29, 115, 0}

\NewDocumentCommand{\heng}
{ mO{} }{\textcolor{red}{\textsuperscript{\textit{Heng}}\textsf{\textbf{\small[#1]}}}}

\NewDocumentCommand{\ef}
{ mO{} }{\textcolor{blue}{\textsuperscript{\textit{Emilio}}\textsf{\textbf{\small[#1]}}}}

\title{Community Moderation and the New Epistemology of Fact Checking on Social Media}
\author[1]{Isabelle Augenstein}
\author[2]{Michiel Bakker}
\author[3,*]{Tanmoy Chakraborty}
\author[4]{David Corney}
\author[5]{Emilio Ferrara}
\author[6]{Iryna Gurevych}
\author[7]{Scott Hale}
\author[8]{Eduard Hovy}
\author[9]{Heng Ji}
\author[10]{Irene Larraz}
\author[11]{Filippo Menczer}
\author[12]{Preslav Nakov}
\author[13]{Paolo Papotti}
\author[12]{Dhruv Sahnan}
\author[1]{Greta Warren}
\author[14]{Giovanni Zagni}

\affil[1]{University of Copenhagen, Nørregade 10, 1172 København, Denmark}
\affil[2]{Massachusetts Institute of Technology, 77 Massachusetts Avenue, Cambridge, MA, United States of America}
\affil[3]{Indian Institute of Technology Delhi, New Delhi, 110016, India}
\affil[4]{Full Fact, 17 Oval Way, London, SE11 5RR, United Kingdom}
\affil[5]{University of Southern California, Los Angeles, CA 90007, USA}
\affil[6]{Technical University of Darmstadt, Hochschulstraße 10, D-64289 Darmstadt, Germany}
\affil[7]{University of Oxford, Broad St, Oxford OX1 3AZ, United Kingdom}
\affil[8]{The University of Melbourne, Grattan Street, Parkville, Victoria 3010, Australia}
\affil[9]{University of Illinois Urbana-Champaign, 506 S. Wright St. Urbana, IL 61801-3633, USA}
\affil[10]{Newtrales, C/Vandergoten 1, 28014 Madrid, Spain}
\affil[11]{Indiana University, 1015 E 11th St., Bloomington, IN 47408, USA}
\affil[12]{Mohamed bin Zayed University of Artificial Intelligence, Masdar City, Abu Dhabi, 7909, United Arab Emirates}
\affil[13]{EURECOM, Campus SophiaTech, 450 Route des Chappes, CS 50193 - 06904 Biot, FRANCE}
\affil[14]{Pagella Politica/Facta, viale Monza 259/265, Milano, 20125, Italy}

\affil[*]{Corresponding author, Email: \href{mailto:tanchak@iitd.ac.in}{tanchak@iitd.ac.in}}

\begin{abstract}
Social media platforms have traditionally relied on internal moderation teams and partnerships with independent fact-checking organizations to identify and flag misleading content. Recently, however, platforms including X (formerly Twitter) and Meta have shifted towards community-driven content moderation by launching their own versions of crowd-sourced fact-checking -- Community Notes. 
If effectively scaled and governed, such crowd-checking initiatives have the potential to combat misinformation with increased scale and speed as successfully as community-driven efforts once did with spam.
Nevertheless, general content moderation, especially for misinformation, is inherently more complex. Public perceptions of truth are often shaped by personal biases, political leanings, and cultural contexts, complicating consensus on what constitutes misleading content.
This suggests that community efforts, while valuable, cannot replace the indispensable role of professional fact-checkers. 
Here we systemically examine the current approaches to misinformation detection across major platforms, explore the emerging role of community-driven moderation, and critically evaluate both the promises and challenges of crowd-checking at scale.

\end{abstract}

\begin{document}

\flushbottom
\maketitle
%
%
\thispagestyle{empty}


\section*{Introduction}

Social media platforms empower users to share opinions and perspectives at scale. This openness brings the persistent challenge of dealing with harmful, misleading, or otherwise objectionable content without unduly constraining freedom of expression. To navigate this tension, platforms implement content moderation policies aimed at protecting users from potentially dangerous material while preserving the integrity of public discourse.
The responsibility of enforcing these rules traditionally falls upon teams of experts such as content moderators, supplemented by veracity judgments provided by third-party independent fact-checkers~\cite{Meta_collab_independent_fcs,tech_coalition_moderator_provider,tiktok_moderation} as well as by automatic Artificial Intelligence (AI) systems working in tandem~\cite{youtube_moderation_guidelines,meta_tech_moderation,tiktok_ai_moderation_report}.
In practice, these platform \emph{enforcers} determine whether to remove content, restrict its visibility, or attach warning disclaimers to a post, depending on its potential to cause harm.
However, both AI-based and centralized manual moderation have limitations.
AI tools for content moderation, while scalable, are marred by high rates of \emph{false positives}, often over-flagging benign content and unfairly reprimanding users, and \emph{false negatives}, letting truly harmful content slip by undetected~\cite{challenges_ai_moderation,auto_content_moderation_increases_adherence,tonneau2024hateday}. 
Recent advances in large language models (LLMs) have shown some promise in automatic veracity prediction~\cite{DeVerna2024LLMFC,zhou2024correctingmisinformationsocialmedia} to support fact-checkers but suffer from issues with factuality~\cite{augenstein2024factuality} and utility in practice~\cite{micallef2022factcheckers,Warren_2025}.
Fact-checking experts are reliable, but cannot keep up with the relentless pace of user-generated content posted online~\cite{nakov2021automatedfactcheckingassistinghuman,can_llms_assist_content_moderation}.

The sheer volume and velocity of online information has strained traditional fact-checking models, leading to innovations in crowd-sourced moderation approaches such as Community Notes \cite{wojcik2022birdwatch}, which aim to leverage crowd wisdom for broader and faster coverage, albeit with its own set of considerations regarding quality and bias.
In an effort to keep content moderation as democratic as possible, the core idea behind this approach is to leverage the collective input of users on the platform (i.e., \emph{community}) to add context to posts that may violate content policies or contain misleading information. 
In this approach, platforms maintain automatic systems and internal teams in place to remove illegal or severely harmful posts containing harassment, violence, sexual exploitation, and drug-related content, while the remainder of posts are subject to community-driven moderation. 
In January 2025, Meta revealed a major policy change, announcing that it would end the use of third-party fact-checkers on its platforms due to alleged bias~\cite{meta_more_speech_fewer_mistakes}.
Meta stated that content that many users might view as acceptable political commentary --- albeit to be consumed with a pinch of salt --- may have been unnecessarily suppressed. This argument is disputed by fact-checkers, who point out that they take serious steps to maintain political impartiality, including regular internal and external reviews \cite{ifcn_open_letter_meta}.

Neither professional fact-checking alone nor nascent community moderation systems like Community Notes offer a perfect solution to online misinformation.
While professional fact-checking offers depth and rigor, its scalability is limited. Conversely, community-driven models like Community Notes promise scalability and diverse perspectives but must navigate challenges of consensus-building and potential manipulation.
Additionally, it builds upon an epistemological proposition that facts are subject to consensus and negotiation, rather than objective or indisputable as traditionally intended by professional fact-checkers. Such a shift has far-reaching potential consequences for the global information eco-system more broadly.  
This paper explores how these distinct approaches have been deployed in content moderation, critically assesses the promises of community moderation, and highlights how collaboration between communities, experts, and technical innovations can address pervasive online misinformation.



\section*{Community-Driven Content Moderation}

The concepts underpinning community-driven content moderation have a long history~\cite{wisdom_of_crowds_book}. 
Wikipedia, for instance, has long operated on the principle of collaborative governance, relying on its users to faithfully curate and manage information for the platform~\cite{wiki_editing}.
Similarly, Reddit delegates some autonomy to its users, allowing a subset of users to moderate content within their respective subreddits~\cite{reddit_mod_coc}. 
Many social media platforms lean on user reports, not only to flag harmful content that evades content moderators, but also to help calibrate policies in response to concerns about certain types of content~\cite{meta_more_speech_fewer_mistakes,yt_policy_development}. However, user reports that inform content moderation decisions implemented by the platforms differ from more structured approaches to community fact-checking. 

In the domain of community-driven content moderation on social media platforms, Twitter (now X) launched \emph{Birdwatch}~\cite{wojcik2022birdwatch} in early 2021, the first large-scale initiative in this space, and later rebranded it \emph{Community Notes}~\cite{twitter_introducing_birdwatch,x_about_community_notes}.
While the program's initial iteration exposed significant shortcomings---such as a vulnerability towards targeted manipulation attempts and partisan bias affecting the notes' writing style and approval, the company has since invested a substantial amount of resources into its refinement~\cite{birds_of_a_feather_rand,x_intro_birdwatch}. 
Key improvements include a more sophisticated algorithm to calculate helpfulness of notes, which rewards notes endorsed by a diverse set of users rather than a simple majority; and eligibility criteria for contributors to ensure participation by genuine users~\cite{x_about_community_notes}.
Following X's lead, other social media giants, such as Meta and, to a lesser extent, TikTok and Weibo, have recently pivoted in favor of a similar community-driven moderation approach over hired experts for moderation on their platforms~\cite{Meta_2025,weibo_announce_cn,tiktok_announce_footnotes}.
Some of these platforms posit community notes as a one-size-fits-all solution to the limitations of fact-checker-led content moderation.
Community notes have indeed shown encouraging results on several fronts where moderation assisted by third party fact-checkers is limited, such as content coverage~\cite{saeed2022crowdsourced,drolsbach2024community,renault_collaboratively_2024}.
However, most of these results and the portrayal of community-driven content moderation as the definitive solution rest on overly-optimistic assumptions about the integrity, diversity, and efficiency of user collaboration.
Just as citizen journalism's initial promise to democratize information 
has faltered and been critiqued for its issues with capacity, reliability and lack of professional standards \cite{mutsvairo2022citizen,splichal2016journalism}, similar limitations may manifest in community notes.

\begin{figure}[t]
    \centering
    \includegraphics[width=0.9\linewidth]{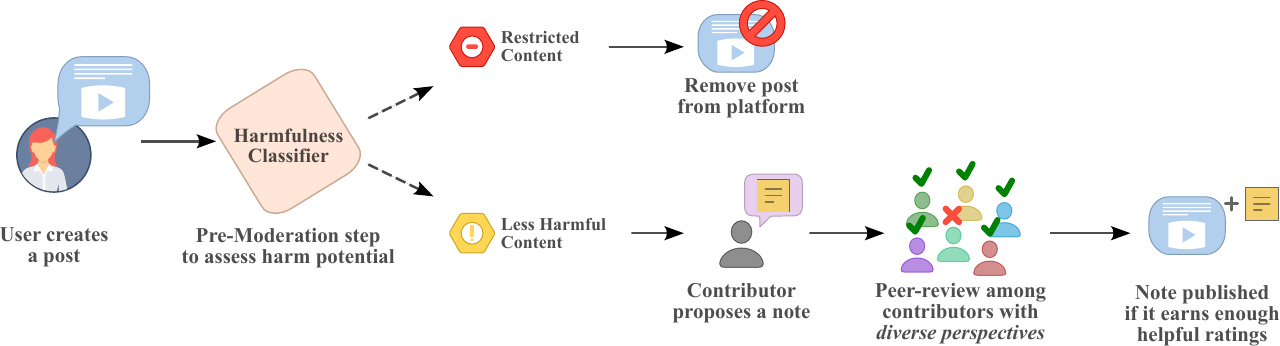}
    \caption{An overview of the community-driven content moderation framework as proposed by major social media platforms like X, Meta, and TikTok. 
    The moderation process is divided into two main stages: \text{(i)} \textbf{Pre-Moderation} using AI classifiers, which categorize content as either \textit{restricted} --- blocked from appearing on the platform --- or \textit{less harmful}, which is passed on for community moderation; and \text{(ii)} \textbf{Community Moderation}, where eligible volunteers may propose additional context that undergoes peer review by other contributors with diverse perspectives before being published after a consensus is achieved.}
    \label{fig:cn_framework}
\end{figure}



In this paper, we review the current implementation of community notes. 
As illustrated in \Cref{fig:cn_framework}, the frameworks deployed by platforms such as X, Meta, and TikTok follow a similar multi-step process~\cite{x_dsa_transparency_report,x_about_community_notes,meta_how_enforce_policies,tiktok_moderation,tiktok_announce_footnotes}. 
All new posts first go through a pre-moderation step, where an automated ``harmfulness'' classifier assesses the harm potential of the content. 
Based on the nature of the content and the inferred type of harm, the content is categorized as either \emph{restricted} or \emph{less harmful}.

Posts with content that poses severe harm, such as calls for violence or terrorism, depictions of child sexual exploitation, and drug-related activities, are deemed to be completely unacceptable by most platforms.
Moreover, the spread of these posts also exposes platforms to expensive lawsuits for failing to protect their users from digital harassment~\cite{x_lawsuit_masjoody,meta_lawsuit_underwood}.
Such content is labeled as \emph{restricted} at the pre-moderation step itself and is not published  on the platforms.
Additionally, the platforms monitor user reports to identify and remove content violations that seep through the automated system. Similarly, content about which a classifier is uncertain may go directly to a queue for a manual check~\cite{tonneau2024hateday}.
Notably, X took down over 3 million posts from public view in the latter half of 2024, either via automated flagging systems or human review~\cite{x_transparency_h2_2024}.

Posts that are not 
automatically 
restricted may still contain misleading or harmful content.
Such posts are
subject to community-driven content moderation, where eligible users may propose \emph{community notes} providing context that clarifies why the content may be wrong or misleading. 
The proposed note, then, undergoes a peer-review process, where other community note contributors with diverse perspectives rate its \emph{helpfulness}.
For a note to appear publicly, it must earn enough helpful ratings that pass the platform's acceptability threshold.
Once rated as \emph{helpful}, the note is displayed 
alongside the original post. While Meta previously prevented posts labeled by independent fact-checkers as false or misleading from being algorithmically promoted, this is not currently the case for posts with community notes.
However, it is important to note that what constitutes `diverse perspectives' remains unclear, as platforms do not clearly define diversity itself -- only how it is quantified in the algorithm using historical mutual disagreement between users on the perceived helpfulness of other community notes. 

\subsection*{The promises of community moderation}

Social media platforms claim that community moderation offers several improvements over expert fact-checking: 

\begin{itemize}[noitemsep]
    \item \emph{\bf The Community Notes model democratizes content moderation.}
Conventionally, fact-checkers advise platforms on potentially harmful mis/disinformation, prompting actions like labeling, limiting spread, or removal. Meta claims this has led to user dissent over perceived censorship~\cite{meta_more_speech_fewer_mistakes}. In contrast, community moderation allows all (non-severely harmful) posts to remain public, letting users decide if a \emph{note of caution} is needed. This may reduce perceived bias and encourage pluralistic interpretations.
However, Wikipedia shows that such systems can be undermined by dominant editors or collusive groups~\cite{wiki_edit_wars}. Community moderation also risks favoring popularity over truth. Moreover, due to echo chambers on social media~\cite{Sasahara_Chen_Peng_Ciampaglia_Flammini_Menczer_2021,arora2022polarization}, community notes may not reach ideologically diverse users, weakening their effectiveness~\cite{martel2023crowdmisinfo}.

\item \emph{\bf Community moderation can scale up and speed up misinformation detection.}
    The volume of user-generated content far exceeds what professional fact-checkers can handle, forcing them to prioritize the most harmful and verifiable claims. Payment models tied to fact-check volume can incentivize checking easier claims, leading to biases and leaving harmful content unchecked. Still, this \textit{selection bias} is arguably justified, as it targets the most dangerous content. Community notes, by contrast, allow a broader pool of volunteers to verify more diverse content. Yet, self-selection remains a challenge -- few have the time or motivation to participate, as seen in the skewed demographics of Wikimedia editors~\cite{Wikireport2018}. Discovery is another issue: due to personalization and polarization, knowledgeable users may not encounter misleading posts~\cite{arora2022polarization}. Moreover, expert verification often comes too late -- most of the engagement typically occurs before a claim is fact-checked~\cite{wack2024politicalfactcheckingeffortsconstrained}. Paid, crowd-sourced moderation can be faster and scale better~\cite{martel2023crowdmisinfo}, but has outperformed experts only in limited cases. Many useful notes remain unpublished due to contributor disagreements, and it is unclear if such systems cover more content overall~\cite{saeed2022crowdsourced,ccdh}.

    \item \emph{\bf Community moderation is less intrusive.}
    Community-driven moderation shows peer-approved notes non-intrusively alongside posts~\cite{meta_more_speech_fewer_mistakes,x_about_community_notes}, allowing users to engage with or ignore them. Earlier expert-led systems often used more intrusive warnings, requiring user action to proceed.
Community notes usually avoid verdicts (e.g., true/false), instead offering missing context -- useful for ambiguous claims or grey areas like dogwhistles~\cite{bhat2020covert}, and less likely to polarize.
However, studies show that expert warning labels effectively reduce belief in and spread of misinformation, even among skeptics~\cite{Martel_Rand_2024,martel_misinfo_warning_labels_2023}.
Ultimately, the `intrusiveness' of moderation is a design choice, and fact-checker systems could adopt similar display styles. Therefore, a direct comparison of the effectiveness of both approaches remains necessary.

\end{itemize}




\section*{Community Notes versus Third-Party Fact-Checking}

\begin{table}[!t]
    \centering
    \resizebox{0.95\textwidth}{!}{
        \begin{tabular}{p{0.15\textwidth}|p{0.4\textwidth}|p{0.4\textwidth}|p{0.4\textwidth}}
        \toprule
           \textbf{Dimension} & \textbf{Third-Party Fact-Checker Based Moderation} & \textbf{Community Notes} & \textbf{Our Critique} \\
           \midrule
           
           Volume 
           & \tcbox[enhanced,box align=base,nobeforeafter,colback=myred!20,colframe=myred,size=small,arc=1.8mm]{Low}   
           
           $\xrightarrow{}$ Can verify only a small fraction of all content.
           & \tcbox[enhanced,box align=base,nobeforeafter,colback=mygreen!20,colframe=mygreen,size=small,arc=1.8mm]{High}
           
           $\xrightarrow{}$ Potential to address a larger volume with thousands of volunteers
           & $\xrightarrow{}$ How well community moderation scales in practice is yet to be proven: only a small proportion of submitted community notes reach publication\\
           \midrule
           
           Breadth 
           & \tcbox[enhanced,box align=base,nobeforeafter,colback=myred!20,colframe=myred,size=small,arc=1.8mm]{Low}
           
           $\xrightarrow{}$ Forced to focus on high-visibility claims given volume of content.
           
           $\xrightarrow{}$ Limited in the scope of issues that can be addressed due to several factors. 
           & \tcbox[enhanced,box align=base,nobeforeafter,colback=mygreen!20,colframe=mygreen,size=small,arc=1.8mm]{High}
           
           $\xrightarrow{}$ Can cover a broader set of topics beyond expert capacity.

           $\xrightarrow{}$ Anonymity can allow crowds to address sensitive claims that experts may be at risk for doing so.
           & $\xrightarrow{}$ Effectiveness is contingent on volunteers willing to engage with claims falling outside of mainstream discourse.\\
           \midrule

           Expertise 
           &  \tcbox[enhanced,box align=base,nobeforeafter,colback=mygreen!20,colframe=mygreen,size=small,arc=1.8mm]{High}
           
           $\xrightarrow{}$ Fact-checkers are trained on the task and can effectively address all sorts of claims.
           & \tcbox[enhanced,box align=base,nobeforeafter,colback=myred!20,colframe=myred,size=small,arc=1.8mm]{Low}
           
           $\xrightarrow{}$ Can re-purpose existing fact-checks to address known misinformation or lower-risk claims.
           & $\xrightarrow{}$ Creation of new knowledge is often required: a skill fact-checkers are trained for, but the crowd is unlikely to exhibit.
           \\
           \midrule

           Speed 
           & \tcbox[enhanced,box align=base,nobeforeafter,colback=myred!20,colframe=myred,size=small,arc=1.8mm]{Slow} 
           
           $\xrightarrow{}$ Rigorous time-intensive quality checks to ensure factual correctness limits the number of claims verified per day.
           & \tcbox[enhanced,box align=base,nobeforeafter,colback=myyellow!20,colframe=myyellow,size=small,arc=1.8mm]{Highly Variable} 
           
           $\xrightarrow{}$ Can be much faster at the verification process by leveraging the ``wisdom of the crowds''.

           & $\xrightarrow{}$ In practice, community notes are faster than fact checks <10\% of the time \cite{saeed2022crowdsourced}.

           $\xrightarrow{}$ Consensus among ideologically diverse users is a critical factor: some notes are approved quickly; other valid ones face delays or are left unpublished.\\
           \midrule

           Democratic Nature 
           & \tcbox[enhanced,box align=base,nobeforeafter,colback=myyellow!20,colframe=myyellow,size=small,arc=1.8mm]{Partly} 
           
           $\xrightarrow{}$ Unilateral decisions by dedicated teams and the platforms to add warning labels / remove content.

           $\xrightarrow{}$ Fact-checker decisions also undergo internal peer-review by fellow fact-checkers.
           & \tcbox[enhanced,box align=base,nobeforeafter,colback=mygreen!20,colframe=mygreen,size=small,arc=1.8mm]{Yes} 
           
           $\xrightarrow{}$ The community decides if a post needs extra context and what that extra context should be.

           $\xrightarrow{}$ Democratic voting process: any user can propose a note, which is made public only if a consensus on its helpfulness is reached.
           & $\xrightarrow{}$ Unclear whether a 'democratic' process contributes to the discovery of the objective truth; may lead to valid notes remaining unpublished
           \\
           \midrule

           Bias 
           & \tcbox[enhanced,box align=base,nobeforeafter,colback=mygreen!20,colframe=mygreen,size=small,arc=1.8mm]{Low} 
           
           $\xrightarrow{}$ Forced to prioritize high-risk claims; introducing a bias in choosing the claims they can verify.

           $\xrightarrow{}$ Selection bias is towards countering harmful content; no evidence of this bias resulting in suppression of authentic narratives barring recent allegations by social media platforms.
           & \tcbox[enhanced,box align=base,nobeforeafter,colback=myyellow!20,colframe=myyellow,size=small,arc=1.8mm]{Moderate} 
           
           $\xrightarrow{}$ No forced prioritization of high-visibility / high-risk claims.
           
           $\xrightarrow{}$ Consensus among users with opposing views is required, making it inherently less biased.

           $\xrightarrow{}$ Crowds are biased towards what they consider interesting to flag and whether the content is even seen by users inclined to flag it.
           & $\xrightarrow{}$ Majority of sources cited as evidence by the crowds are left-leaning in practice.
           
           $\xrightarrow{}$ Crowds are also susceptible to various cognitive biases, while fact-checkers are trained to view content through a neutral lens.
           \\
           \midrule

           Transparency 
           & \tcbox[enhanced,box align=base,nobeforeafter,colback=myyellow!20,colframe=myyellow,size=small,arc=1.8mm]{Moderate} 
           
           $\xrightarrow{}$ Platforms offer limited justification for censorship of content, thereby making it harder to scrutinize expert decisions.
           
           $\xrightarrow{}$ Third-party fact-checkers are bound by strict principles of transparency of bias, sources and funding
           & \tcbox[enhanced,box align=base,nobeforeafter,colback=myyellow!20,colframe=myyellow,size=small,arc=1.8mm]{Moderate}

           $\xrightarrow{}$ Deliberation process can be fully transparent, with users being able to view all proposed notes and the level of agreement within the community on these notes.

           & $\xrightarrow{}$ The algorithm lacks key details, especially how diversity among users is defined. 
           
           $\xrightarrow{}$ Unclear how users may scrutinize the community's decisions. 

           $\xrightarrow{}$ Lack of transparency regarding bias and conflicts of interest of contributors
           
           $\xrightarrow{}$ No onus on the crowd to outline the verification process; fact-checkers generally present this as part of their analysis of the claim.
           \\

           \midrule

           Coordinated\newline Adversarial\newline Attacks 
           & \tcbox[enhanced,box align=base,nobeforeafter,colback=mygreen!20,colframe=mygreen,size=small,arc=1.8mm]{Safe} 

           $\xrightarrow{}$ No easy way for malicious actors to attack the sanctity of the professional fact-checking process.
           & \tcbox[enhanced,box align=base,nobeforeafter,colback=myred!20,colframe=myred,size=small,arc=1.8mm]{Vulnerable} 
           
           $\xrightarrow{}$ Susceptible to coordinated attacks that challenge the credibility of authentic information, or suppress helpful notes due to dissenting views. 
           & $\xrightarrow{}$ Ample evidence of malicious users colluding to inorganically boost content on social media.
           
           $\xrightarrow{}$ Wikipedia exhibits vulnerability to collusion in the voting system for approving specific edits on the website.
           \\
           \midrule

           Impact on misinformation beliefs and spread
           & \tcbox[enhanced,box align=base,nobeforeafter,colback=mygreen!20,colframe=mygreen,size=small,arc=1.8mm]{High} 

           $\xrightarrow{}$ Fact-checker labels are effective in preventing misinformation, even among users skeptical of fact-checkers.
           & \tcbox[enhanced,box align=base,nobeforeafter,colback=myyellow!20,colframe=myyellow,size=small,arc=1.8mm]{Mixed}

           $\xrightarrow{}$ Found effective in preventing spread of misinformation in some cases.

           $\xrightarrow{}$ Evidence suggests it expedites voluntary retraction of misleading posts.
           
           $\xrightarrow{}$ Considerable time delay between publication of the post and the community note.

           & $\xrightarrow{}$ Notes are not ``promoted'' by the platform in any way. People who have already seen the post do not go back and read the note.
           
           $\xrightarrow{}$ Fact-checkers publicise their work, communicate retractions and even directly approach public figures to ask for corrections.
           \\ 
           \midrule

           Psychological Harm  
           & \tcbox[enhanced,box align=base,nobeforeafter,colback=myyellow!20,colframe=myyellow,size=small,arc=1.8mm]{Argued} 

           $\xrightarrow{}$ High cognitive load in viewing sensitive content, but protective policies are in place for welfare of experts.

           $\xrightarrow{}$ Fact-checkers are professionally trained and accustomed to encountering sensitive content.
           & \tcbox[enhanced,box align=base,nobeforeafter,colback=myyellow!20,colframe=myyellow,size=small,arc=1.8mm]{Not Discussed}

           $\xrightarrow{}$ No protective policies proposed for the cognitive load experienced in viewing sensitive content.
           & $\xrightarrow{}$ Psychological risks associated with laypeople encountering sensitive content are left undiscussed by social media platforms.

           $\xrightarrow{}$ Lack of protective policies exposes platforms to legal liabilities.

           $\xrightarrow{}$ That said, rowd-sourced moderation is voluntary, and users are not forced to consume/verify sensitive content.
           \\
           \midrule
            
           Unpaid Labour
           & \tcbox[enhanced,box align=base,nobeforeafter,colback=mygreen!20,colframe=mygreen,size=small,arc=1.8mm]{No}

           $\xrightarrow{}$ Third-party fact-checkers are hired by platforms to advise on handling of misleading content.
           & \tcbox[enhanced,box align=base,nobeforeafter,colback=myred!20,colframe=myred,size=small,arc=1.8mm]{Yes}

           $\xrightarrow{}$ Relies on unpaid volunteers to engage in the task without labor protection.
           & $\xrightarrow{}$ Addressing misleading content is a demanding task, which deserves adequate remuneration.

           $\xrightarrow{}$ Shifting responsibility to unpaid volunteers diminishes the fact-checkers' work and undermines the public's right to ``the whole truth''.
           \\
           \bottomrule
        \end{tabular}
    }
    \caption{Comparison of capabilities of \textit{Third-party Fact-checker based Moderation} and \textit{Community Notes} on social media platforms.
    Our analysis uses a three-point scale: Low capability in red, Mixed or Unclear capability in yellow, and High capability in green.
    We also outline the proposed advantages and drawbacks of both moderation approaches as argued by key stakeholders (e.g., social media platforms and fact-checkers), along with brief details on our critical evaluation of these claims.}
    \label{tab:comparison_exp_comm}
\end{table}

The rhetoric and policy shifts by social media platforms such as Meta and X suggest that community notes are a salve to the issues with fact-checking~\cite{meta_more_speech_fewer_mistakes,twitter_introducing_birdwatch}.
We believe this is a false dichotomy: the two approaches are deeply intertwined~\cite{borenstein2025communitynotesreplaceprofessional}. 
Here, we present a balanced investigation of the extent to which community notes delivers on the claimed benefits. We shed light on potential risks for users and platform integrity, which make community notes ill-suited as a comprehensive replacement of moderation experts. 
\Cref{tab:comparison_exp_comm} presents a brief overview of our discussions in subsequent subsections, offering a comparative breakdown, across various indicators, of the two moderation approaches: use of third-party fact-checkers and reliance on community moderation.

\subsection*{Are community notes more scalable?}

By harnessing the ``wisdom of the crowd,'' the community notes model has a strong potential to enhance the scalability and quicken the response times of content moderation on social media platforms\cite{martel2023crowdmisinfo}.
However, analyses of similar crowd-sourced approaches demonstrate competing evidence that cast doubt on the claims of increased effectiveness over expert-led moderation.


\begin{itemize}[noitemsep]
    \item \emph{\bf Volume:} Community notes can be written and voted on by any platform user (subject to a minimum quality check\cite{x_writingnotes}), while professional fact-checkers are typically limited to individuals with training in journalism. 
    Given the limited capacity of fact-checking projects (as of May 2024, there were 439 independent professional fact-checking projects in 111 countries \cite{Duke2024FactCheckingSputters}), 
    community notes hold the potential to address a much larger volume of misleading claims than fact-checkers alone. However, given that just a small proportion of proposed notes reach publication status~\cite{saeed2022crowdsourced}, 
    the volume of moderated posts does also not correspond to the (output) volume of warning labels. 

    \item \emph{\bf Breadth:} 
Fact-checkers outside Western, Educated, Industrialised, Rich, and Democratic (WEIRD) nations face challenges like limited press freedom~\cite{balod2021fighting}, data scarcity~\cite{cheruiyot2018fact}, financial constraints~\cite{ababakirov2022meeting,Warren_2025}, and even physical threats~\cite{vinhas2023weird}, despite support from global networks~\cite{GlobalFactCheckFund2025}.
Community notes can extend coverage to regions where professional fact-checking is constrained. Pseudonymity allows laypeople and citizen journalists to flag misinformation in risky environments.
Professionals prioritize high-virality or high-harm claims~\cite{Neumann2022misinfoharms,nakov2021clef}, whereas community notes can address a wider range—provided volunteers are present. However, contributor demographics (e.g., Wikipedia) remain skewed toward Western males~\cite{Wikireport2018,graham2015digital}.
Only 20\% of contributors have written notes that reach consensus\cite{wirtschafter2023future}, indicating that self-selection may limit both participation and topic diversity.

    \item \emph{\bf Expertise:} 
    Fact-checking is a skilled and complex task that requires specific experience, expertise, and data access, for example, in identifying patterns of misinformation and uncovering large-scale disinformation campaigns.
    Furthermore, verifying complex or high-stakes claims (e.g., relating to health, science, or economics) often requires specific expertise, which most community note-writers are unlikely to possess.
    Often, the information needed to verify a claim is not available online and requires creating new knowledge, e.g., by directly contacting experts or first-hand witnesses \cite{Warren_2025,graves2017anatomy}.
    In these cases, crowd-sourced fact-checking tends to rely on existing analyses by professional fact-checkers \cite{borenstein2025communitynotesreplaceprofessional,Zhao_Naaman_2023}. 
    Community note writers often target lower-risk misleading posts, such as scams \cite{borenstein2025communitynotesreplaceprofessional} and claims that have previously been fact-checked \cite{saeed2022crowdsourced}. 
    

    \item \emph{\bf Speed:} 
While professional fact-checking articles undergo lengthy cross-checks and editorial oversight to ensure quality, community notes have been posited as a method of expediting the fact-checking process. However, there is mixed evidence in support of this claim:  one study found 20–30 ideologically diverse laypeople can reach accurate verdicts faster than experts~\cite{martel2023crowdmisinfo}, but another showed the `crowd' was quicker than experts in less than 6\% of cases~\cite{saeed2022crowdsourced}.
A key bottleneck is the requirement for cross-ideological agreement before publishing a note, leading to an average delay of 15.5 hours~\cite{renault_collaboratively_2024} and many helpful notes never being published~\cite{ccdh}.
The system may also be vulnerable to manipulation by coordinated actors suppressing valid notes~\cite{Mahadevan_2025}.
In collaborative models like Wikipedia-style editing, laypeople were often faster than experts only because they could reference existing professional fact-checks~\cite{Zhao_Naaman_2023}, indicating the complementarity between expert and volunteer fact-checking.
    
\end{itemize}

\subsection*{Are community notes more trustworthy?}

One of the areas in which community notes hold the greatest promise is their potential to increase public trust in fact-checking, content moderation, and social media platforms. 
However, there are several significant barriers that must be addressed to realize this potential.

\begin{itemize}[noitemsep]
    \item \emph{\bf Democratic:} 
    The current implementation of Community Notes on X stipulates that in order to be published, a note must receive enough `helpful' votes from users who generally tend to disagree with one another in their votes. 
    This method rewards writing notes that readers with distinct perspectives can agree upon, and ensures that notes are not monopolized by certain ideologies.
    Notes that refer to \textit{trustworthy} \cite{prollochs_community-based_2022} or \textit{unbiased}\cite{solovev2025referencesunbiasedsourcesincrease} sources tend to result in higher `helpfulness' ratings.
    However, due to political polarization \cite{yasseri2023crowdsourcing}, community notes on contentious political issues rarely reach a consensus\cite{saeed2022crowdsourced,ccdh}.
    This approach can therefore quash fact-checks on politically-sensitive claims, even if there is a clear and indisputable verdict.
    We note a core epistemological difference in how the two approaches view facts: community notes as subjective constructs that are a matter of personal opinion, compared to fact-checking as an objective truth to be discovered through analyzing evidence \cite{uscinski2013epistemology,amazeen2015revisiting}. 
    This raises a fundamental question for platforms, policymakers, and the public: can democratic legitimacy and epistemic truth coexist in content moderation?


    \item \emph{\bf Bias:} Proponents of community notes have argued that the cross-perspective consensus required for publication means that these crowd-sourced fact-checks are inherently less politically biased than those written by professional fact-checkers.
    However, the majority of sources that are cited in notes are left-leaning news outlets rated to be of high factuality\cite{kangur2024checkscheckersexploringsource}, 
    suggesting that similar perspectives emerge in fact-checking and community notes.
    Moreover, notes for posts from US Republicans are proposed more often and rated as more helpful than for posts from US Democrats \cite{renault_mosleh_rand_2025}. 
    Community note users may also be susceptible to various cognitive biases: crowdworkers generally overestimate the truthfulness of claims, are overconfident in their abilities to judge the truthfulness of statements, and their truthfulness judgments are adversely impacted by their opinion of the claimant\cite{draws_effects_2022}.
    It has also been highlighted that while professional fact-checks are expected to adhere to formal and neutral communication styles, there is no such onus on community volunteers. Note writers may attempt to manipulate readers using highly persuasive but logically incoherent argumentation\cite{kankham_community_2024}.

 \item \emph{\bf Transparency:} 
   Platforms like X and Meta have open-sourced their community notes algorithms~\cite{xcommnote,Meta_2025}, enabling external inspection and reuse. In principle, this promotes transparency, letting users and readers understand how notes are assigned.
While professional fact-checkers present evidence logically, their decision-making often involves intuition~\cite{Warren_2025,graves2017anatomy}, as likely with community reviewers. Still, algorithmic transparency may boost trust.
In practice, however, documentation lacks detail. For instance, ideological diversity is modeled along a single axis—adequate for political bias~\cite{wojcik2022birdwatch}, but insufficient for capturing cultural, linguistic, or conspiratorial biases, requiring further study.
Additionally, note authors remain anonymous, obscuring potential biases that could be inferred from past behavior. In contrast, fact-checking organizations follow strict transparency standards~\cite{IFCNPrinciples2016,EFCSNStandards2022}, often naming individual fact-checkers and disclosing sources, funding, and correction policies.
Community notes lack a formal correction system; updates depend on users to detect and vote on inaccuracies.

    \item \emph{\bf Coordinated adversarial attacks:} 
    Online social networks have long been susceptible to coordinated adversarial attacks --- be it from automated bots or groups of users who inorganically inflate social reputations, amplify specific narratives and engage in other potentially harmful behaviors~\cite{Pacheco2021Coordinated,bradshaw2017troops}.
    Recent studies emphasize that community moderation is also exposed to the same vulnerability~\cite{saeed2022crowdsourced,martel2023crowdmisinfo}.
    Although platforms implement safeguards to validate contributors as real people and not adversarial actors, these mechanisms are not perfect and can be gamed with the right resources.
    Thus, entrusting a virtually unrestricted user base with content moderation responsibilities opens the door for malicious groups to exploit the system.
    By creating an artificial perception of internal disagreement through collusive inorganic activity on benign content, such groups can deceive the community notes algorithm and gain the platform's trust.
    This could place them in a dangerously advantageous position, allowing them to launch coordinated attacks on authentic narratives --- suppressing their reach by casting doubt on their credibility through misleading community notes that falsely debunk the original content.
    
\end{itemize}





\subsection*{Do community notes help in countering misinformation?} 

There is insufficient data on the effectiveness of community notes as interventions for misinformation.

\begin{itemize}[noitemsep]
    \item \emph{\bf Impacts on misinformation beliefs:} The additional context provided by community notes appears to be appreciated by readers: they are judged as more trustworthy than simple labels that flag misinformation with no additional detail\cite{drolsbach2024community}.
    However, the study did not compare community notes with context written by professional fact-checkers, so it remains unclear whether a clear preference exists between the two approaches. 
    Community notes and news article suggestions were found to be equally effective in reducing people's belief in and intention to repost misleading posts on social media\cite{kankham_community_2024}. 
    An analysis of a single health-related claim in the same study found that community notes were more effective interventions for a positive framing (e.g., ``This food may cure cancer''), whereas related articles were more effective for a negative framing (e.g., ``This food may cause cancer''). 
    It is unclear how far this may generalize to other claims and domains. 
    Notes may also have negative tradeoffs: displaying community notes leads users to post more negative, angry, disgusted replies to misleading posts\cite{chuai_community_2024-1}.

    \item \emph{\bf Impacts on misinformation spread:} Although Meta and X do not automatically reduce the reach or visibility of a post that has received a note \cite{meta_more_speech_fewer_mistakes}, there is some evidence suggesting that notes are useful in curbing misinformation spread: posts on Twitter/X that are labeled by community notes as `misleading' receive 37\% fewer retweets than posts deemed `not misleading'.~\cite{drolsbach_diffusion_2023} 
    Community notes also increase the probability of tweet retractions and deletions and expedite the retraction of misleading posts.~\cite{gao_can_2024,renault_collaboratively_2024} 
    On the other hand, community notes can draw attention to misleading posts, leading to increases in likes, engagement, and followers for accounts that receive `misleading' community note labels.~\cite{wirtschafter2023future}

\end{itemize}

\subsection*{Are community notes an ethical replacement for professional fact-checkers?}

Several ethical concerns have been raised with regard to an increasing reliance by social media platforms on crowd-sourced fact-checking.

\begin{itemize}[noitemsep]
    \item \emph{\bf Psychological harm:} Professional fact-checkers are accustomed to encountering harmful content (e.g., violence, abuse, and other explicit content) regularly, and receive dedicated training and support to manage the psychological consequences. 
    Shifting this work to volunteers without any form of psychological support risks severe harm to those who write and rate community notes \cite{steiger2021psychological}. 
    There is also a risk for platforms of potential liability and legal action by content moderators\cite{MIT3reasons2025}.

    \item \emph{\bf Unpaid labor:} In addition to the emotional burden borne by community moderators, relying solely on community notes raises the ethical issue of expecting non-professional volunteers to carry out intricate research on behalf of for-profit companies (distinct from contributors to non-profit Wikipedia) without remuneration or labor protection.
    The task of fact-checking requires skilled, challenging work, and those who undertake it should be fairly compensated for it. 

    \item \emph{\bf De-funding and de-professionalization of fact-checking:} 
    In addition to eroding fact-checking organizations' capacity to perform their basic functions, divestment in fact-checking devalues and diminishes the critical importance of access to reliable information for functioning societies~\cite{moran2025endofTS}. 
    Although community moderation relies on unpaid labor, it lacks professional standards, particularly in terms of training, accountability, and methodology.
    Community notes should be viewed as an approach that complements professional fact-checking rather than replacing it.

\end{itemize}

\section*{Recommendations}

\begin{table}[t]
    \centering
    \resizebox{\textwidth}{!}{
        \begin{tabular}{p{0.13\textwidth}|p{0.2\textwidth}p{0.2\textwidth}p{0.4\textwidth}}
        \toprule
           \textbf{Category} & \textbf{Recommendation} & \textbf{Challenges Addressed} & \textbf{Details}\\
           \midrule
           Collaboration between community and experts & Workload distribution & \vspace{-7.5pt}
           \begin{itemize}[noitemsep, nolistsep, topsep=0pt,leftmargin=1em]
               \item Breadth
               \item Volume
               \item Expertise
           \end{itemize}
           \vspace{-7.5pt}& \vspace{-7.5pt}
           \begin{itemize}[noitemsep, nolistsep, topsep=0pt,leftmargin=1em]
               \item Crowdwork to verify repetitive misinformation or widely debunked claims with ref
               \item Fact-checkers to concentrate on new high-risk claims that need creation of new knowledge
           \end{itemize}
           \vspace{-7.5pt}\\
           \cmidrule{2-4}
           & Fact-checkers as secondary reviewers & \vspace{-7.5pt}
           \begin{itemize}[noitemsep, nolistsep, topsep=0pt,leftmargin=1em]
               \item Bias
               \item Speed
           \end{itemize}
           \vspace{-7.5pt} & \vspace{-7.5pt}
           \begin{itemize}[noitemsep, nolistsep, topsep=0pt,leftmargin=1em]
               \item Fact-checkers can assess notes subjectively, considering all sides of the narrative
               \item Fact-checkers to act as reviewers and approve notes that show partial helpfulness but do not meet the platform's threshold
           \end{itemize}
           \vspace{-7.5pt} \\
           
           \cmidrule{2-4}
           & Flagging investigation-worthy claims & \vspace{-7.5pt}
           \begin{itemize}[noitemsep, nolistsep, topsep=0pt,leftmargin=1em]
               \item Volume
               \item Expertise
               \item Bias
           \end{itemize}
           \vspace{-7.5pt} & \vspace{-7.5pt}
           \begin{itemize}[noitemsep, nolistsep, topsep=0pt,leftmargin=1em]
               \item Use community notes to identify and/or prioritize check-worthy claims for fact-checkers
               \item Transparency through providing overview of distinct ideological groups writing and flagging posts
           \end{itemize}
           \vspace{-7.5pt} \\
           \midrule
           Collaboration between technology and the community & Social opinion analytics & \vspace{-7.5pt}
           \begin{itemize}[noitemsep, nolistsep, topsep=0pt,leftmargin=1em]
               \item Speed
               \item Adversarial attacks
           \end{itemize}
           \vspace{-7.5pt}
           & \vspace{-7.5pt}
           \begin{itemize}[noitemsep, nolistsep, topsep=0pt,leftmargin=1em]
               \item Expedite bridging diverse perspectives by identifying users more likely to engage in constructive debate
               \item Improve robustness to brigading and/or coordinated attacks
           \end{itemize}
           \vspace{-7.5pt} \\
           \cmidrule{2-4}
           & Fusing Community Notes & \vspace{-7.5pt}
           \begin{itemize}[noitemsep, nolistsep, topsep=0pt,leftmargin=1em]
               \item Bias
           \end{itemize}
           \vspace{-7.5pt}
           & \vspace{-7.5pt}
           \begin{itemize}[noitemsep, nolistsep, topsep=0pt,leftmargin=1em]
               \item Identify points of agreement and discord between users
               \item Generate notes that diverse perspectives can agree on
           \end{itemize}
           \vspace{-7.5pt} \\
           \cmidrule{2-4}
            & AI-agents to simulate crowds & \vspace{-7.5pt}
           \begin{itemize}[noitemsep, nolistsep, topsep=0pt,leftmargin=1em]
               \item Bias
               \item Volume
           \end{itemize}
           \vspace{-7.5pt}
           & \vspace{-7.5pt}
           \begin{itemize}[noitemsep, nolistsep, topsep=0pt,leftmargin=1em]
               \item Flag posts with potentially sensitive content for review by experts
           \end{itemize}
           \vspace{-7.5pt} \\
           \cmidrule{2-4}
           & Previously ``community-noted posts'' & \vspace{-7.5pt}
           \begin{itemize}[noitemsep, nolistsep, topsep=0pt,leftmargin=1em]
               \item Volume
           \end{itemize}
           \vspace{-7.5pt}
           & \vspace{-7.5pt}
           \begin{itemize}[noitemsep, nolistsep, topsep=0pt,leftmargin=1em]
            \item Recommend notes from related posts
            \item Enable cross-platform community moderation
           \end{itemize}
           \vspace{-7.5pt} \\
           \bottomrule
        \end{tabular}
    }
    \caption{A summary of our recommendations to address the challenges faced by community notes.
    We present the list of our recommendations, the specific challenges they address, and a brief description of how they could be implemented within the community moderation algorithm.}
    \label{tab:recommendations}
\end{table}

Drawing from our discussions thus far, it is clear that community moderation presents several unresolved issues that keep it from realizing its potential. 
Social media platforms have not yet addressed these issues when endorsing the community-driven approach as a replacement for expert-led moderation.
We argue that 
platforms must improve the design of community moderation algorithms, taking into account their assumptions regarding the efficient user collaboration that is necessary for the success of this moderation approach. 
Here we put forth some recommendations for how the current issues with community moderation can be addressed by platforms and policymakers.
\Cref{tab:recommendations} provides a summary of our discussion in this section.

\subsection*{Collaboration between community and experts}


Community moderation represents an important step towards democratizing and scaling up content moderation. 
Yet we believe that its adoption as a replacement for fact-checkers is a missed opportunity for fruitful collaboration between experts and the crowd. 
Several reputed experts have advocated for such a collaborative approach~\cite{poynter_angie_2025,MIT3reasons2025,ifcn_open_letter_meta}.
Involving fact-checkers at various steps in the community moderation framework can overcome many of its unresolved issues:

\begin{itemize}[noitemsep]
    \item \emph{\bf Distributing the workload by filtering claims based on risk and ease of verification:} The crowd can address ``low-hanging fruit'' of mis- and disinformation. 
    Previously, third-party fact-checking programs for content moderation have used machine learning solutions to address low-risk content, but with minimal human oversight.
    Community-driven content moderation allows volunteers to propose notes on relevant content by leveraging previously fact-checked claims and AI to surface already verified information---a technique that has proven effective in the mitigation of misleading content~\cite{gao_can_2024}.
    This, in turn, would allow professional fact-checkers to focus on emergent, high-risk claims that demand deeper investigation with an expert touch.

    \item \emph{\bf Fact-checkers acting as secondary reviewers to approve valid notes stuck in peer-review:} 
    A collaborative approach can help alleviate bias-related issues and resolve cases where helpful notes do not reach consensus.
    Fact-checking experts can provide an expert, third-party assessment of the notes and content, analyzing both sides of the story where disputing opinions among laypersons might stall the peer review process.
    Earlier, some platforms had in-house teams of moderators who would take action against potentially harmful content, based on the advise of third-party fact-checkers.
    Now with the community notes model, platforms can, similarly, appoint further layer of governance comprising of established fact-checking experts.
    Through an internal content moderation management system, platforms can empower these experts to act as reviewers who can grant approval to helpful notes proposed by volunteers.
    This could be particularly useful for valid notes that exhibit some agreement within the community, but fall short of the acceptability threshold due to partisan objections. We note that research has experimented with using AI to help with this task.\cite{de2025supernotes}

    
    \item \emph{\bf Flagging investigation worthy claims for professional fact-checkers:}
    The crowd could flag content in need of deeper analysis by expert fact-checkers. 
    Fact-checkers often spend significant time just looking for claims that present the greatest potential risk for society. 
    Having the crowd's support would help professional fact-checkers filter the vast stream of user-generated content to a smaller set of relatively high-risk claims. 
    Platforms have always presented user-reported content to hired independent fact-checkers for verification.
    Community moderation presents a more transparent and democratic way of implementing such a flagging or priority system, though the concerns raised about transparency above still apply. 
    This system could further be augmented with a transparent view of the users engaging with investigation-worthy content. Such contextual analysis, commonly included in expert fact-checks, helps illuminate patterns in how misinformation spreads and which groups are most affected.

    \item \emph{\bf Cross-platform community notes:} 
    Part of the success in removing child sexual abuse material (CSAM) and other clearly illegal material is the existence of centralized, third-party repositories (e.g., The Internet Watch Foundation and GIFCT). No such resource linking fact-checks or community notes to content currently exists.
    One of the major strengths of X's Community Notes approach is its open source algorithm, which has been adopted by Meta \cite{Meta_2025}. 
    A centralized community moderation system could be facilitated by multiple social media platforms, so notes that appear on one platform appear for the same claim on others. Such an approach could also help platforms with fewer users respond faster to misleading or other harmful content.
    
    
\end{itemize}


\subsection*{Collaboration between technology and the community}

Alongside the collaboration between fact-checking experts and the crowd, we propose how technology could address some of the pitfalls of community moderation. 
AI and network analysis can improve the efficiency of key stages of the current community moderation, and even automate certain processes---ultimately enhancing the productivity of the crowd.


\begin{itemize}[noitemsep]
    \item \emph{\bf Auditing `diverse' perspectives:}
    As we have discussed, the lack of transparency with regard to the users of community notes remains a challenge with potential for technical solutions.
    For example, proposing quantitative metrics to measure the diversity of perspectives, or developing AI techniques for improving the diversity of and de-biasing such clusters.
    
    \item \emph{\bf Using network analytics to address the partisan bottlenecks:}
    A major source of tension and debate among experts is how the algorithm handles the bridging of diverse perspectives in the peer review process.
    Reaching a consensus in community moderation on certain issues is not as easy as the platforms make it sound, evident from how valid notes are often left unpublished due to insufficient votes~\cite{Mahadevan_2025,ccdh}.
    Research on modeling social opinion dynamics offers network analytics methods to efficiently identify users with opposing perspectives by studying their online activity~\cite{yasseri2023crowdsourcing,Sasahara_Chen_Peng_Ciampaglia_Flammini_Menczer_2021}.
    Some studies even analyze how users with competing views could be selected from the network such that they would engage in a constructive debate on a particular topic~\cite{reducing_controversy_connecting_opposing,factors_recc_contrarian_cont}. 
    Such methods could help address the major bottleneck of community moderation by finding people who are willing to rate notes that potentially oppose their own views. 

    \item \emph{\bf Using network analytics to identify collusive groups:}
    The community notes algorithm is at risk from collusive behavior by malicious user groups.
    In particular, these groups may: \emph{(i)} deceive the algorithm through manufactured internal disagreement to launch attacks on authentic content at opportune moments; or \emph{(ii)} suppress helpful notes by rating them negatively simply because they conflict with the group's preferred narrative.
    Prior research on social network analysis has proposed frameworks for detecting collusive behavior on social media -- such as groups of users artificially inflating social reputations or amplifying specific narratives~\cite{dutta2020blackmarketdrivencollusiononlinemedia}.
    Although early detection of artificial disagreement among community notes contributors is a fundamentally different task compared to existing efforts, we believe these frameworks could still offer useful foundations for spotting coordinated manipulation in community moderation.
    Further research in this sphere is critical.
    Researchers and platforms must continue analyzing community notes data to uncover collusive activities and the underlying patterns of user interaction to build robust defenses against such manipulation in community moderation.

    \item \emph{\bf Using AI to fuse information from various proposed community notes:}
    As discussed earlier, a significant number of valid notes remain unpublished because users with opposing viewpoints often fail to reach consensus on them~\cite{ccdh}.
    Some studies suggest that this could partly be because the proposed notes present a biased narrative -- one that aligns with the contributor's perspective, sometimes omitting key information.
    Moreover, this also causes ``the whole truth'' to be fragmented across multiple notes, with no single note providing a clarity on the situation.
    Consequently, users from diverse perspectives find it hard to rate a particular note as helpful.
    To address this, \emph{Supernotes} have been proposed, which are AI-generated notes that integrate information from all proposed notes for a post into a single, cohesive version~\cite{de2025supernotes}.
    These \emph{Supernotes} are designed to maximize the probability of consensus among users by reflecting historical patterns of writing in community notes rated as helpful by diverse user groups~\cite{de2025supernotes}.
    We believe that \textit{Supernotes} and other similar methods hold great promise for community moderation by synthesizing notes that offer a comprehensive and neutral account of events -- making them more acceptable across a broad range of user perspectives.

    \item \emph{\bf AI agent-augmented crowds for proposing community notes:}
    With recent advancements in reasoning capabilities of LLMs, agentic fact-checking has shown great promise for verifying real-world-claims~\cite{}.
    Parallelly, as community notes have become prominent, some studies have also advocated for the potential of LLM-agents to simulate crowds for community moderation~\cite{costabile2025assessingpotentialgenerativeagents}.
    These swarms of agents have demonstrated an ability to classify the truthfulness of social media posts, with performance comparable to that of human crowds.
    However, it is important to keep in mind that these studies were performed in controlled environments, but real-world, user-generated data is much more noisy and complex.
    Further research is needed to evaluate the performance of such systems in practical, real-time scenarios.
    That said and building on insights from prior efforts in developing successful agentic fact-checking systems, we believe that LLM agents could be useful for proposing notes on problematic content.
    And as community notes grow, these agents could be further tuned to align with the performance of ideal human crowd-workers, particularly to overcome common limitations of human crowds---such as truthfulness overestimation and cognitive biases~\cite{draws_effects_2022}.
    Note that, we do not advocate for fully replacing human community notes contributors.
    Rather, we envision these AI-agents to augment human efforts in proposing community notes by automating certain parts of the process.

    \item \emph{\bf Using AI to suggest similar previously `community-noted' posts:}
    A key step in automatic fact-checking is to identify whether a seemingly new claim can be verified using a previously fact-checked claim~\cite{nakov2021clef,shaar-etal-2020-known}.
    A similar strategy could be adopted for community notes.
    Such a system could also be augmented to enable \emph{cross-platform community moderation}.
    Given the transparent and open-source nature of community notes, platforms adopting it could collaborate to centralize these notes.
    AI could then match new posts to previously ``community-noted'' posts, minimizing redundant verification efforts from the community across platforms.


    
\end{itemize}




\section*{Conclusion}

Community notes offer significant promise for addressing misinformation on social media platforms, with the potential to increase the speed, scale, and participation in fact-checking.
Yet our analysis of the current implementations of the community note model reveals critical limitations of crowdsourced fact-checking and demonstrates that they fall well short of a comprehensive solution to the socio-technical challenges of misinformation, and to the complex issues of bias and trust in information.
In this paper, we have outlined how some of these challenges might be addressed.
Collaborating with fact-checkers and journalists can complement a community moderation system and provide the necessary expertise and accuracy crucial in complex and high-stakes contexts. 
Technical approaches can support the transparency of this approach and bolster resilience to bias and malign manipulation.
Our recommendations for the future of community notes point to how such a model can scale the reach and participation in fact-checking. 
Community-driven efforts are necessary, but not sufficient alone as a means of combating misinformation.
Social media platforms must take responsibility for foregrounding and incentivizing high-quality contributions, while policymakers can mandate transparency and common standards among platforms, to ensure that systems are designed to serve the broader information society.
Community moderation may help democratize fact checking, but without the integration of expert viewpoints, algorithmic transparency, and institutional support, it risks consolidating consensus over establishing correctness.


\section*{Author contributions statement}

Conceptualized by I.A., T.C. and P.N.; initial manuscript prepared by D.S., G.W., I.A., T.C, P.N.; all authors contributed to finalizing the manuscript. The author names are arranged alphabetically by last name.



\bibliography{references}



\end{document}